# High-throughput Screening of Quaternary Compounds and New Insight for Excellent Thermoelectric Performance

Aijun Hong[*#a], Yuxia Tang[#b], and Junming Liu[*b]

[a]*Jiangxi Key Laboratory of Nanomaterials and Sensors, School of Physics, Communication and Electronics, Jiangxi Normal University, Nanchang 330022, China*

[b]*Laboratory of Solid State Microstructures, Nanjing University, Nanjing 210093, China*

[*]Correspondence and requests for materials should be addressed to A.J.H. (E-mail: 6312886haj@163.com and liujm@nju.edu.cn).

[#]Aijun Hong and Yuxia Tang contribute equally to the article.

**Abstract:** It is well known that the high electric conductivity, large Seebeck coefficient, and low thermal conductivity are preferred for enhancing thermoelectric performance, but unfortunately, these properties are strongly inter-correlated with no rational scenario for their efficient decoupling. This big dilemma for thermoelectric research appeals for alternative strategic solutions, while the high-throughput screening is one of them. In this work, we start from total 3136 real electronic structures of the huge $X_2YZM_4$ quaternary compound family and perform the high-throughput searching in terms of enhanced thermoelectric properties. The comprehensive data-mining allows an evaluation of the electronic and phonon characteristics of those promising thermoelectric materials. More importantly, a new insight that the enhanced thermoelectric performance benefits substantially from the coexisting quasi-Dirac and heavy fermions plus strong optical-acoustic phonon hybridization, is proposed. This work provides a clear guidance to theoretical screening and experimental realization and thus towards development of performance-excellent thermoelectric materials.

**Introduction**

Thermoelectric (TE) materials have been driving extensive concern due to the specific functionalities of transferring waste heat into electrical energy and vice versa.[1] Their performance can be quantified by dimensionless parameter $ZT = S^2\sigma T/\kappa$, namely the figure of merit, where $S$ is the Seebeck coefficient, $\sigma$ the electrical conductivity, $T$ the absolute temperature, and $\kappa$ the total thermal conductivity containing carrier contribution $\kappa_e$ and phonon contribution $\kappa_L$. Obviously, a good TE material should possess high power factor ($PF = S^2\sigma$), low $\kappa$, and thus high $ZT$.[2-7] However, a realization of high $ZT$ and $PF$ becomes intrinsically challenging since this set of physical parameters ($S$, $\sigma$, $\kappa$) exhibit complex interdependences or say, are intrinsically coupled.[8-10] For instance, an improved $\sigma$ or reduced $\kappa_L$ is often accompanied by the reduction of $S$ or rise of $\kappa_e$, making an independent modulation of these parameters almost impossible but yet highly unremitting.

The unremitting effort continues to focus on the electronic structure of TE materials in order to find alternative scenario for decoupling these interdependences. On one hand, Chasmar and Stratton[11] first defined the material parameter $B$ to evaluate the TE performance,

$$B = T\frac{k_B^2 h}{3\pi^2}\frac{c_l N_{v,c}}{m^* \lambda^2 \kappa_L} \tag{1}$$

where $k_B$ and $h$ are the Boltzmann constant and Plank constant, $c_l$ and $\lambda$ stand for the average longitudinal elastic constant and deformation potential constant, respectively, $N_v$ and $N_c$ define the numbers of degenerate bands for valence and conduction bands, $m^*$ is the band effective mass for conduction carriers. One can see that the excellent TE performance (high $ZT$) requires large $N_{v,c}$ and small band effective mass $m^*$.[12-14] Subsequently, Liu et al[3] proposed the generalized material parameter $B^*$, where a large weighted mobility $U^*$ and a large bandgap $E_g$ are preferred to result in high $ZT$ value. In short, according to these theories, the electronic structure of an excellent TE material should be featured with small $m^*$, large $N_{v,c}$, and large $E_g$, simultaneously.

Certainly, this scenario is questionable and an unavoidable fact is that the most excellent TE materials are of $p$-type carriers with relatively small bandgap $E_g$ (< 1.0 eV).[15] The underlying physics is that the $p$-type systems possess large density of states (DOS) effective mass $m_D = N_v^{2/3} m^*$ due to the large $m^*$, benefiting to enhanced Seebeck coefficient $S$, according

to the Pisarenko formula in the degenerate limit:[13, 16-17]

$$S = \frac{8\pi^2 k_B^2 T}{3eh^2}(\frac{\pi}{3n})^{\frac{2}{3}} m_D, \qquad (2)$$

where $e$ and $n$ are the elementary charge and carrier density, noting that a small $E_g$ facilitates large $n$ and thus realization of high $\sigma$. It seems that a reciprocal relation between large $S$ and high $\sigma$ becomes essentially inevitable.

Nevertheless, recent experiments did reveal that the *p*-type FeNbSb-based half-Heusler (HH) alloys show the hole mass $m^* \sim 2.0\ m_e$ ($m_e$ for the free electron mass) along the Γ-L direction and $E_g \sim 0.54$ eV. The obtained $ZT$ maximum, $ZTM$, may reach up to $\sim 1.1$ at temperature $T \sim 1100$ K.[18-19] This implies that the reciprocal relation between $S$ and $\sigma$ may be avoided and a high TE performance is possible without following the rule of small $m^*$, large $N_{v,c}$, and large $E_g$ simultaneously, likely suggesting extra-opportunities to realize high $\sigma$ and large $S$ beyond the reciprocal relation between $S$ and $\sigma$, although the window may be narrow. Based on this perspective, searching for promising TE candidates from the whole family of binary and ternary compounds has been continued for decades, indeed discovering some high-performance TE materials. Nevertheless, for most cases investigated earlier, the $S \sim \sigma$ reciprocal relation remains unbroken, and the reason is more or less related to the fact that these compounds have limited degrees of freedom for electronic structure variations. A natural strategy is to find opportunity in the family of more-element compounds which have wider variety of crystal and electronic structure degrees of freedom. Along this line, the quaternary compound family has been occasionally touched for multi-motivations such as superconductivity,[20] thermoelectricity[21] and topological insulator,[22] however, no systematic investigation on the TE properties of the whole family has been available.

So far available experimental data suggested that the quaternary compounds possessing general chemical formula $X_2YZM_4$, where X, Y, and Z are most likely transition metal species and M = Se, Te, may be promising. In particular, $Cu_2FeSnSe_4$[21, 23] and $Cu_2ZnSnM_4$ (M=S, Se)[24] were reported to have good TE properties, hinting that this family could be a rich bank. In addition, this family exhibits in many cases relatively low thermal conductivity as a concerned issue, due to the strong atomic anharmonic vibration,[25] evidenced by the thermal conductivity as low as $\sim 0.31$ Wm$^{-1}$K$^{-1}$ at a medium temperature for compound $Ag_2BaSnSe_4$[26] in comparison

with HH alloys that show high $\kappa_L$.[27-30]

This $X_2YZM_4$ family often crystallizes in tetragonal phase with space groups (SG) such as I-42m and I-4, trigonal phase (SG: $P3_1$, $P3_2$ and R3c), and orthorhombic phase (SG: Ama2 and I222).[31-33] They have rich polytypism, and different structures may show very different behaviors in terms of the TE properties. For example, $Cu_2ZnSnSe_4$ are TE-favored in the stannite structure (SG: I-42m),[24] while they show outstanding optical absorption in the Kesterite structure (SG: I-4) or the coexisting I-42m and I-4 structures.[34-36] Fortunately, a close check of the whole $X_2YZM_4$ family finds that the lattice symmetry can be roughly categorized into these four dominant template phases: I222, Ama2, I-42m, and $P3_1$, as shown in Figure 1a-d. Certainly, the number of compounds with the four dominant phases can be thousands.

Given these opportunities and major issues, a high-throughput strategy for identifying these $X_2YZM_4$ compounds is highly appealed. In this work, we target this strategy by addressing two major issues within the proposed framework. On the one hand, a thorough screening of the whole family and thus a determination of the energy-minimal lattice structure for each will be carried out, focusing on the four dominant phases that belong to space groups I222, Ama2, I-42m, and $P3_1$. On the other hand, a thorough checking of the electronic structures of those compounds will be performed, from which a new insight connecting the structure ~ TE property relationship will be proposed. Eventually, we have considered the total 3136 structures and their TE properties, and found a number of promising TE compounds that have not been reported earlier. Based on this database, we are able to establish a close connection between the high TE performance and a novel scenario that the coexisting quasi-Dirac and heavy fermions (CDHF) in combination with strong optical-acoustic phonon hybridization (OAPH) constitutes a new framework for discovering excellent TE materials, at least for this $X_2YZM_4$ family but clearly extendable to other compounds.

**Method**

The high-throughput first-principles calculations are carried out in the Vienna Ab-initio Simulation Package (VASP) using the generalized gradient approximation (GGA) with the Perdew-Burke-Ernzerhof functional (PBE).[37] For the structure optimizations, we choose different $k$ meshes for various SG types: 8×8×8 for I222, and I-42m, 8×8×4 for Ama2 and $P3_1$.

The total number of *k* points expands to eight times in the self-consistent and high-precision calculations. The PBE with the modification of Becke-Johnson (mBJ)[38] calculations are performed within the full-potential code WIEN2k[39] for obtaining the exact energy eigenvalues and then TE parameters are calculated by solving electronic Boltzmann transport equation (EBTE), details of which were reported previously.[40] Noted that，we choose 15×15×15 *k* meshes for I222, and I-42m, 15×15×7 *k* meshes for Ama2 and P31 in the mBJ calculations. We combine the compressive sensing lattice dynamics (CSLD) method[41-42] and the phonon Boltzmann transport equation (PBTE),[43] and calculate the phonon dispersion relationship and the lattice thermal conductivity.

## Results & Discussion

### Crystal structure and PBE-bandgap distribution

In this computational package, all the 3136 compounds in the $X_2YZM_4$ family, including several polytypic structures for the identical chemical composition in some cases, are submitted to full geometry optimization. Such optimized structure database allows us to further evaluate the electronic properties on one hand, and on the other hand a close connection between the electronic structure and lattice geometry features such as polyhedrons. Indeed, a thorough checking suggests the critical roles played by the polyhedrons in determining the band structure. Our previous work[25] reported that the valence band (VB) and conductivity band (CB) edges of compounds $Ag_2YZSe_4$ (Y = Ba, Sr; Z = Sn, Ge) with I222 symmetry mainly come from the contributions of polyhedrons [$XM_4$] and [$ZM_4$], respectively.

Along this line, it is found that most of the I222, Ama2, and P3$_1$ phases have three types of staggered [$XM_4$], [$YM_8$], and [$ZM_4$] polyhedrons, while [$YM_8$] changes to [$YM_4$] in the I-42m phase (in Figure 1e-g). Certainly, in some cases these polyhedrons may appear in unfixed forms upon different elements and lattice constants even for the same compound with different polytypic units. For example, [$YM_4$] may appear in the I222, Ama2, and P3$_1$ phases for $Cu_2CdHfSe_4$ while [$YM_8$] exists in most I-42m-type $Ag_2Ba$-based compounds. Usually, for a $X_2YZM_4$ compound, the major contributions to the VB and CB edges comes from at least two elements, which allows more degrees of freedom for electronic structure variations, and thus more opportunities to engineer the band structure. One can regulate the VB shape without much variation of the CB by modifying [XM4], and the CB shape by modifying [ZM4]. Therefore, it is of significance to map the intrinsic dependences of the electronic structure on the polyhedron feature or space group (lattice symmetry). In addition, chemical doping or defect engineering on these polyhedrons [XM4] and [ZM4] can effectively tune the electronic structure because the latter is dependent of the electronegativity differences between X and M and/or between Z and M.

The primary feature for electronic structure is its bandgap $E_g$. The first target for the present work is to evaluate the relationship between the bandgap and lattice features (space group, polyhedron form etc). While details of the database on bandgap are listed in Table S1 of the Supplemental Materials (SM), we present the histogram patterns regarding the number of compounds within various $E_g$ ranges, in Figure 2a-d for the Te-based compounds and in Figure 2e-h for the Se-based compounds, noting M = Te and Se. It is seen that the Te-based compounds are mainly narrow bandgap semiconductors ($E_g$ < 1.0 eV), of which there are totally only 48 compounds with wide bandgap ($E_g$ > 2.0 eV) that are commonly inapplicable as TE materials. Differently, the most Se-based compounds have their bandgap distribution around ~ 2.0 eV, due to the stronger electronegativity of element Se, leading to the large electronegativity difference between (X, Y, Z) and M. The number of highly insulating Se-based compounds reaches up to 298, implying that it is difficult for these insulators to be chosen for charge doping in order to achieve high TE performance.

Here, it should be mentioned that a choice of proper bandgap for a good TE material remains to be an issue in many cases. For those narrow bandgap semiconductors, the underlying physics is given by the generalized material parameter $B^*$ in expression of $U^*$, $\kappa_L$, and $E_g$:

$$B^* = 6.668 \times 10^{-2} \frac{U^*}{\kappa_L} E_g, \qquad (3)$$

suggesting that a relatively large $E_g$ is preferred for a high $B^*$ and high $ZT$. This is reasonable because the carriers near valence band maximum (VBM) and conductivity band minimum (CBM) both participate in electronic transport if $E_g$ is too small, leading to the bipolar effect suppressing Seebeck coefficient $S$ in spite of enhancing conductivity $\sigma$ concomitantly. Thus, it seems that those compounds with $E_g$ ~ 1.0 eV would be highly preferred.

The data shown in Figure 2 suggest that most of the Te-based compounds can be categorized into this class as parent compounds for subsequent optimization via carrier doping and band engineering, while this is just a preliminary screening based on the bandgap requirement. For those compounds with $E_g \sim$ [2.0 eV, 3.0 eV], the optimal carrier density $n_{opt}$ required for realizing the high $ZT$ generally corresponds to the Fermi level $E_f$ located at the CBM for $n$-type carriers or the VBM for $p$-type carriers. This implies a change of $E_f$ for more than 1.0 eV is needed for band engineering via doping, nano-crystallization, and defects. It becomes thus inevitable to deform the electronic structure topology that usually hinders the improvement of TE performance. In this sense, most of the Se-based compounds would be inapplicable as TE materials. Even so, if a bandgap of 1.0 eV, as obtained here from the PBE-scheme, is chosen as the primary criterion, there are totally 1538 structures that can be categorized into this class. These compounds would be submitted to subsequent procedure of screening.

**mBJ corrected bandgaps and *ZTM* screening**

It is known that a calculation using the PBE scheme always under-estimates the bandgap, and thus a more accurate scheme of calculation is needed. The mBJ scheme is more costly but can give much more accurate $E_g$ value. Consequently, the total 1538 structures are calculated using the mBJ scheme and the obtained $E_g$ values are used for subsequent TE property evaluation in the semi-classical Boltzmann transport theory. In this work, a variant of $ZT$ factor[44-45] is adopted to sort the TE performance from the calculated electronic structure:

$$ZT = \frac{S^2 \frac{\sigma}{\tau} T}{\frac{\kappa_e}{\tau} + \frac{\kappa_L}{\tau}}, \qquad (4)$$

where $\tau$ is the relaxation time for carriers. Given the assumption of constant relaxation time, ratios $\sigma/\tau$ and $\kappa_e/\tau$ can be directly obtained by solving the electron Boltzmann transport equation (EBTE).[40] In our calculations, we took $\tau = 3.33 \times 10^{-15}$ s by keeping the conservative factor $\kappa_L/\tau = 3 \times 10^{14}$ Wm$^{-1}$K$^{-1}$s$^{-1}$.

It is well known that the $ZT$ factor is strongly dependent of temperature $T$ and carrier density $n$. For each of the selected compounds, one chooses three temperatures ($T$ = 300, 600, and 900 K), six carrier densities ($n = 5 \times 10^{17}$, $5 \times 10^{18}$, $5 \times 10^{19}$, $5 \times 10^{20}$, $1 \times 10^{21}$, and $5 \times 10^{21}$ in units of cm$^{-3}$), and two carrier types (*p*-type and *n*-type) to calculate the *ZT* factor, counting totally 36 *ZT* values for each compound. Then the maximal of each 18 *ZT* values is denoted as 0.*ZTM* for each carrier type and the data are collected in Table S2 of the Supplemental Materials. Furthermore, in order to check the effectiveness of the screening method, we compare the calculated results and some available experimental data on compound Cu$_2$SnZnSe$_4$-based alloys. While our prediction of the *ZTM* for stannite Cu$_2$SnZnSe$_4$ is 0.97 at *T* = 900 K, the measured *ZT* value was 0.91 at *T* = 860 K,[46] confirming the nice consistence between them. The detailed discussion is shown in Tables S3 and S4 the Supplemental Materials.

All the evaluated data below $E_g \sim 2.0$ eV are mapped into the totally 16 ($E_g$, *ZTM*) plots, as shown in Figure 3a-p where the results are categorized by (1) the type of carriers: *n*-type and *p*-type; (2) the lattice symmetry or SG: I222, Ama2, I-42m, and P3$_1$; and (3) the bandgap feature: direct and indirect bandgaps, respectively. In each plot, all the data are fitted using the linear regression of *ZTM* as a function of $E_g$. While the data look somehow strongly scattered, some major features from these plots can be highlighted. First, a clarification of the influence of carrier type suggests that the *p*-type compounds are basically better than those *n*-type compounds. This clarification is basically consistent with the well-known experimental facts. It also suggests that the present screening strategy works well in a qualitative sense. For a quantitative estimation, one focuses on the compounds in one group, such as Te-based compounds of I222 phase. It is seen that 56 *p*-type compounds have their *ZTM* larger than one, while this number for the *n*-type compounds is only 3. Second, no remarkable preference of the direct-bandgap compounds or indirect-bandgap ones is identified, suggesting that this band feature is not an essential ingredient for TE performance although it could be the core one for photovoltaic or optoelectronic applications. Indeed, sufficient data from earlier experiments support this fact and here one does not need to care it. Third, referred to Eq.(3) where parameter $B^*$ is shown to be a positively linear function of $E_g$, factor *ZTM* as a function of $E_g$ would behave similarly. Indeed, such linear-like behavior can be claimed, as shown by the blue solid lines in Figure 3 for a guide of eyes. However, the data are seriously scattered and no technically sound conclusion can be reached, suggesting that Eq. (3) is at most an insufficient description of these data. Furthermore, such behaviors, if applicable, are not always positively linear, and for the structures of P3$_1$ symmetry and *p*-type carriers, one sees the negatively linear behaviors (Figure

3h,p), which is unusual.

In short, one is in a good position to claim that for these $X_2YZM_4$ quaternary compounds, the TE properties seem to deviate remarkably from the well-established scenario. The well-known SPB model likely exhibits some drawback in describing the electronic and phonon transports. A key issue here is that these compounds have their electronic structure seriously deviating from the parabolic band character. An apparent updating of the current transport theory may be needed, which could be one of the core issues for predicting high-performance TE materials from these quaternary compounds.

**Coexisting quasi-Dirac and heavy fermions (CDHF) scenario**

As discussed earlier, this quaternary compound family includes complex chemical compositions and lattice structures, offering ingredients well beyond the current scenarios on TE materials. While various models that may levitate one physical ingredient more than others can be proposed, it is natural to argue that the underlying physics could contain all these physical ingredients that are not fully included in these models.

Before going into detailed discussion, one may highlight the landscapes of current TE theories. As back to 1950s, Ioffe motivated an idea that guided profoundly the modern TE transport theories. Slack summarized the ideas in 1995 and proposed the famous phonon-glass electron-crystal concept.[47] This concept outlines the macroscopic requirements for electrical and thermal transports of an excellent TE material, while no unambiguous physical scenario on the microscopic electronic and phonon structures has been established. As a rough classification, one may note that the Dirac fermions[48] and heavy fermions[49] seem to take the two complementary ends in constructing the TE performance. For a better illustration, one may consider the band structure of a semiconductor with the heavy fermion character or the Dirac fermion character, and the two types of band structures are drawn in Figure 4a for a guide of eyes. Clearly, there exists a flat band around some high-symmetry point for a heavy fermion system, while the band exhibits nearly linear dispersion for a Dirac fermion system. It is known that a Dirac system usually possesses zero effective mass and small bandgap, thus a high $\sigma$ is easily accessible but a large $S$ becomes impossible. On the other hand, a heavy fermion system usually has large $m^*$ and $S$ for a given $E_g$, while sufficiently high $\sigma$ may not be possible even if a sufficiently high carrier density is achieved. In addition, it has been believed that a multi-valley band structure[50] may benefit to the improved TE performance, evidenced by TE materials such as SnSe,[51] CoSb$_3$[13, 52-53] and PbTe.[54] These scenarios allow us to combine them into one composite framework so that the TE performance in such quaternary compounds can be promoted mutually. Here it should be mentioned that a perfect Dirac fermion transport is characterized by the ideal gapless band structure with perfect linear dispersions on both the conduction band and valence band. Such systems are rare and may not satisfy simultaneously all the requirements. Instead, we consider those gapped systems that exhibit approximately linear dispersion on either the valence band or the conduction band. Therefore, the so-called Dirac fermion band here is more accurately called the quasi-Dirac band.

This strategy is indeed supported by our calculations, and the quaternary compounds with coexisting quasi-Dirac and heavy fermion (CDHF) characters do exist, as evidenced by the data shown in Figure 4b,f. We call these compounds the CDHF-like systems whose quasi-Dirac fermions and heavy fermions respectively support the high $\sigma$ and large $S$ respectively, and reasonably high TE performance can be expected, at least the high power factor ($PF$) can be obtained by proper decoupling of $S$ and $\sigma$. Based on a screening of all the lattice and electronic structures in number of thousands, we determine the energy-lowest crystal phases and then calculate their TE properties. Those compounds with large $ZT$ values are selected into a sub-category for careful identification of the electronic structure in terms of the quasi-Dirac fermion or heavy fermion characters. While it is non-realistic for every electronic structure in this sub-category to have clear CDHF characters, one finds that many of them do favor the coexistence of the quasi-Dirac fermion and heavy fermion configurations. We plot the electronic band structures of selected TE materials with excellent performances in Figure S1. It is found that conductivity band of the n-type and valley band of the p-type exhibits CDHF characteristic.

As representative examples, we show in Figure 4b-e the electronic structure of Cu$_2$ZnHfSe$_4$ with the I-42m symmetry and its TE properties, and in Figure 4f-i the band structure of Cs$_2$BeZrSe$_4$ with the Ama2 symmetry and its TE properties. In the two-band structures (Figure 4b,f), the heavy-fermion band (blue line) and Dirac-like band (red line) are marked for better view. It is immediately seen that the two compounds exhibit remarkable CDHF characters, and in fact, they are the best *n*-type and *p*-type TE materials found in this work. Their $ZT$ values reach 1.26 (for *n*-type $Cu_2ZnHfSe_4$) and 2.05 (for *p*-type $Cs_2BeZrSe_4$) at $T \sim 900$ K, respectively, owing to the high $\sigma$ and especially large $S$ for $Cu_2ZnHfSe_4$ ($S \sim -227.8$ μVK$^{-1}$) and $Cs_2BeZrSe_4$ ($S \sim 296.4$ μVK$^{-1}$). The *PF* values can be $\sim 27.5$ μWcm$^{-1}$K$^{-2}$ and ultrahigh 46.1 μWcm$^{-1}$K$^{-2}$, comparable with $PF \sim 45$ μWcm$^{-1}$K$^{-2}$ at $T = 1100$ K for the *p*-type FeNbSb-based HH alloys [18]. Besides, the best compounds in the Te-based compounds are the *n*-type $Ag_2ZnHfTe_4$ with the I-42m symmetry ($ZT = 1.11$) and the *p*-type $Cs_2BeZrTe_4$ with the Ama2 symmetry ($ZT = 1.69$), as shown in Table S5 of the Supplemental Materials. The slightly large *ZT* values of the *n*-type $Cu_2ZnHfSe_4$ and *p*-type $Cs_2BeZrSe_4$ with respect to the two Te-based compounds are owing to their relatively low $\kappa_e$.

It should be mentioned that the TE parameters ($S$, $\sigma$ and *PF*) for *n*-type $Cu_2ZnHfSe_4$ and *p*-type $Cs_2BeZrSe_4$ are sensitive to the Fermi level location inside the bandgap, owing to the two-band (VB and CB) effect. Obviously, there are two peaks for Seebeck coefficient $S$ but only one single valley for conductivity $\sigma$. The extremes and locations of the *S*-peaks are determined by the bandgap size and edge structure, and the peak locations are more important than the extreme magnitude because the magnitude is usually sufficient for the high performance requirements. Unfortunately, for most TE materials reported so far, the large extreme values imply the low $\sigma$ due to the coupled $S$ and $\sigma$, making *PF* small and a decoupling of them highly concerned.

In the present CDHF scenario, such a decoupling becomes possible by a shifting of the *S*- and/or $\sigma$-features so that a large $S$ and a high $\sigma$ can be balanced simultaneously to reach the optimized *PF*. To be specific, the *n*-type and *p*-type *S* feature locations can be shifted towards the vicinities of VB and CB respectively. It looks like "one small shifting for the *S*- and/or $\sigma$-features contributes to one giant leap for the TE community". Indeed, the *S* features of the *n*-type $Cu_2ZnHfSe_4$ and *p*-type $Cs_2BeZrSe_4$ are biased towards the CB and VB respectively. Consequently, the two systems have their respective *PF* to be tuned up to the maximal, as shown in Figure 4e,i. It also evidences that a TE compound with the CDHF character in the band structure does allow the decoupling of $S$ and $\sigma$, a substantial consequence of the present study.

**Optical-acoustic phonon hybridization**

Nevertheless, such an effective decoupling strategy does likely optimize the power factor

(*PF*) but may not necessarily guarantee the realization of large *ZT* value since the thermal conductivity has so far not been discussed. For the two compounds, we also calculate the phonon spectrum for checking the dynamic stability of the lattice structures (I-42m for $Cu_2ZnHfSe_4$ and Ama2 for $Cs_2BeZrSe_4$) and their thermal conductivity. The spectra are summarized in Figure 5a,e respectively. First, it is seen that there are no imaginary modes in the spectra, indicating that the structures are dynamically stable. Second, one sees clearly the existence of the strong hybridization between the transverse optical (TO) modes and longitudinal acoustic (LA) modes in the two compounds, resulting in remarkable avoided crossing effect.

This avoided crossing effect can easily result in steep variation of the TO and LA modes simultaneously and even their softening. It is seen that the lowest-lying optical (LLO) mode has its frequency as low as 0.31 THz (= 1.3 meV) at the Γ point for $Cs_2BeZrSe_4$, and no behaviors appear for $Cu_2ZnHfSe_4$. This mode crossing-free or softening or both can be an important ingredient for the lattice thermal conduction suppression by enhancing the phonon scattering rate, thus leading to the strong low-frequency anharmonic vibration.

Motivated by the above novel physics, one can combine the compressive sensing lattice dynamics (CSLD) method [41-42] and the phonon Boltzmann transport equation (PBTE) [43], and calculate the lattice thermal conductivity $\kappa_L$ as a function of *T* for $Cu_2ZnHfSe_4$ and $Cs_2BeZrSe_4$, as shown in Figure 5d,h respectively. As expected, $Cs_2BeZrSe_4$ has a lower $\kappa_L$ than that of $Cu_2ZnHfSe_4$ over the whole *T*-range, and $\kappa_L \sim 0.64$ $Wm^{-1}K^{-1}$ at *T* = 300 K and it falls down to 0.28 $Wm^{-1}K^{-1}$ at *T* = 900 K, noting that the well-known TE compound SnSe has the similar $\kappa_L$.

For more detailed analysis of the thermal transport, one sees that the low-frequency phonon anharmonic vibration can be measured by its Grüneisen coefficient $\gamma$. The frequency-dependent $\gamma$ plots show that many modes have large negative $\gamma$ near the low-frequency phonons that have important contributions to $\kappa_L$, implying that these modes can reduce the $\kappa_L$. The more negative the $\gamma$, the smaller the $\kappa_L$. For $Cs_2BeZrSe_4$, the largest |$\gamma$| reaches ~ 39 (see Figure 5g) near the zero frequency, indicating the strong anharmonic vibration due to the avoided crossing and optical mode softening. Here, in spite of more negative Grüneisen coefficients for $Cu_2ZnHfSe_4$, the largest |$\gamma$| is ~ 16 (see Figure 5c) and most of these negative modes are far from the zero frequency. Therefore, $Cu_2ZnHfSe_4$ still has relatively higher $\kappa_L$ than $Cs_2BeZrSe_4$.

**Conclusions**

By utilizing the high-throughput first-principles calculations we have investigated the electronic band structures and TE properties of the huge number of quaternary compounds with general chemical formula $X_2YZM_4$ where M = Te and Se. A structure database consisting of 3136 compounds with four seminal space groups (I222, Ama2, I-42m, and $P3_1$) by the high-throughput screening calculations, has been constructed. Subsequently, we have performed more accurate band structure calculations on the properly chosen 1441 compounds using the mBJ scheme and thus evaluated their TE properties using the combined first-principles electronic and phonon computations and Boltzmann theory. A systematic screening of the electronic and phonon structures of those predicted excellent TE compounds suggests that the coexisting Dirac heavy fermions (CDHF) plus the strong optical-acoustic phonon hybridization (OAPH) appear to be the essential ingredient of physics for remarkably enhanced TE performance, and this scenario would be helpful for experimental and theoretical design of promising TE materials. Finally, tens of promising $X_2YZM_4$ quaternary compounds are predicted for experimental realizations.

**Supporting Information**

The Supporting Information is available free of charge on the ACS Publications website at DOI:xxxxx.

The sizes of PBE-bandgap and mBJ-bandgap, PBE-bandgap type (indirect/direct), total energy per atom, TE parameters of $X_2YZM_4$ compounds in the four phases (I222, Ama2, I-42m and P31).

**Acknowledgements**

Authors gratefully acknowledge great support from the National Natural Science Foundation of China (Grant Nos. 11804132 and 51721001).

**Author contributions**
A. J. H and J. M. L conceived and designed the project. A. J. H performed ab initio calculations. All authors contributed to the writing and editing of the manuscript.

**Competing interests**

The authors declare no competing interests.

1. Aseginolaza, U.; Bianco, R.; Monacelli, L.; Paulatto, L.; Calandra, M.; Mauri, F.; Bergara, A.; Errea, I., Phonon Collapse and Second-Order Phase Transition in Thermoelectric Snse. *Phys Rev Lett* **2019**, *122*, 075901-075901.
2. He, W. K., et al., High Thermoelectric Performance in Low-Cost $Sns_{0.91}se_{0.09}$ Crystals. *Science* **2019**, *365*, 1418-1418.
3. Liu, W. S.; Zhou, J. W.; Jie, Q.; Li, Y.; Kim, H. S.; Bao, J. M.; Chen, G.; Ren, Z. F., New Insight into the Material Parameter B to Understand the Enhanced Thermoelectric Performance of $Mg_2sn_{1-x-y}ge_xsb_y$. *Energy Environ Sci* **2016**, *9*, 530-539.
4. Pei, Y. Z.; Wang, H.; Snyder, G. J., Band Engineering of Thermoelectric Materials. *Adv Mater* **2012**, *24*, 6125-6135.
5. Bell, L. E., Cooling, Heating, Generating Power, and Recovering Waste Heat with Thermoelectric Systems. *Science* **2008**, *321*, 1457-1461.
6. He, J. G.; Amsler, M.; Xia, Y.; Naghavi, S. S.; Hegde, V. I.; Hao, S. Q.; Goedecker, S.; Ozolins, V.; Wolverton, C., Ultralow Thermal Conductivity in Full Heusler Semiconductors. *Phys Rev Lett* **2016**, *117*, 046602.
7. Hong, A. J.; Ma, L. L., Ultralow Thermal Conductivity in Quaternary Compound $Ag_2basnse_4$ Due to Square-Cylinder Cage-Like Structure with Rattling Vibration. *Appl Phys Lett* **2021**, *118*, 143903.
8. Feng, Z.; Fu, Y.; Putatunda, A.; Zhang, Y.; Singh, D. J., Electronic Structure as a Guide in Screening for Potential Thermoelectrics: Demonstration for Half-Heusler Compounds. *Phys Rev B* **2019**, *100*, 085202-085202.
9. Wang, C.; Chen, Y. B.; Yao, S. H.; Zhou, J., Low Lattice Thermal Conductivity and High Thermoelectric Figure of Merit in $Na_2mgsn$. *Phys Rev B* **2019**, *99*, 024310-024310.
10. Zhang, X.; Pei, Y., Manipulation of Charge Transport in Thermoelectrics. *npj Quantum Materials* **2017**, *2*, 68.
11. Chasmar, R. P.; Stratton, R., The Thermoelectric Figure of Merit and Its Relation to Thermoelectric Generators. *Journal of Electronics and Control* **1959**, *7*, 52-72.
12. Hong, A. J.; Gong, J. J.; Li, L.; Yan, Z. B.; Ren, Z. F.; Liu, J. M., Predicting High Thermoelectric Performance of Abx Ternary Compounds Namgx (X = P, Sb, as) with Weak Electron-Phonon Coupling and Strong Bonding Anharmonicity. *J Mater Chem C* **2016**, *4*, 3281-3289.
13. Tang, Y. L.; Gibbs, Z. M.; Agapito, L. A.; Li, G.; Kim, H. S.; Nardelli, M. B.; Curtarolo, S.; Snyder, G. J., Convergence of Multi-Valley Bands as the Electronic Origin of High Thermoelectric Performance in $Cosb_3$ Skutterudites. *Nat Mater* **2015**, *14*, 1223-1228.
14. Wang, H.; Pei, Y. Z.; LaLonde, A. D.; Snyder, G. J., Weak Electron-Phonon Coupling Contributing to High Thermoelectric Performance in N-Type Pbse. *Proc Natl Acad Sci U S A* **2012**, *109*, 9705-9709.
15. Yang, L.; Chen, Z. G.; Dargusch, M. S.; Zou, J., High Performance Thermoelectric Materials: Progress and Their Qpplications. *Adv Energy Mater* **2018**, *8*, 1701797-1701797.
16. Zhao, L. D.; He, J. Q.; Berardan, D.; Lin, Y. H.; Li, J. F.; Nan, C. W.; Dragoe, N., Bicuseo Oxyselenides: New Promising Thermoelectric Materials. *Energy Environ Sci* **2014**, *7*, 2900-2924.
17. Chen, Z. G.; Han, G.; Yang, L.; Cheng, L. N.; Zou, J., Nanostructured Thermoelectric Materials: Current Research and Future Challenge. *Progress in Natural Science-Materials International* **2012**, *22*, 535-549.
18. Fu, C. G.; Zhu, T. J.; Liu, Y. T.; Xie, H. H.; Zhao, X. B., Band Engineering of High Performance P-Type Fenbsb Based Half-Heusler Thermoelectric Materials for Figure of Merit Zt > 1. *Energy Environ Sci* **2015**, *8*, 216-220.
19. Fu, C. G.; Bai, S. Q.; Liu, Y. T.; Tang, Y. S.; Chen, L. D.; Zhao, X. B.; Zhu, T. J., Realizing High Figure of Merit in Heavy-Band P-Type Half-Heusler Thermoelectric Materials. *Nat Commu* **2015**, *6*, 8144-8144.
20. Vlasko-Vlasov, V. K.; Welp, U.; Koshelev, A. E.; Smylie, M.; Bao, J. K.; Chung, D. Y.; Kanatzidis, M. G.; Kwok, W. K., Cooperative Response of Magnetism and Superconductivity in the Magnetic Superconductor $Rbeufe_4as_4$. *Phys Rev B* **2020**, *101*, 104504-104504.
21. Song, Q., et al., Enhanced Carrier Mobility and Thermoelectric Performance in $Cu_2fesnse_4$ Diamond-Like

Compound Via Manipulating the Intrinsic Lattice Defects. *Materials Today Physics* **2018**, *7*, 45-53.

22. Banerjee, K.; Son, J.; Deorani, P.; Ren, P.; Wang, L.; Yang, H., Defect-Induced Negative Magnetoresistance and Surface State Robustness in the Topological Insulator Bisbtese$_2$. *Phys Rev B* **2014**, *90*, 235427-235427.

23. Song, Q. F.; Qiu, P. F.; Zhao, K. P.; Deng, T. T.; Shi, X.; Chen, L. D., Crystal Structure and Thermoelectric Properties of Cu$_2$fe$_{1-x}$mn$_x$snse$_4$ Diamond-Like Chalcogenides. *Acs Applied Energy Materials* **2020**, *3*, 2137-2146.

24. Shi, X. Y.; Huang, F. Q.; Liu, M. L.; Chen, L. D., Thermoelectric Properties of Tetrahedrally Bonded Wide-Gap Stannite Compounds Cu$_2$znsn$_{1-x}$in$_x$se$_4$. *Appl Phys Lett* **2009**, *94*, 122103-122103.

25. Hong, A. J.; Yuan, C. L.; Liu, J. M., Quaternary Compounds Ag$_2$xyse$_4$ (X = Ba, Sr; Y = Sn, Ge) as Novel Potential Thermoelectric Materials. *J Phys D Appl Phys* **2020**, *53*, 115302-115302.

26. Kuo, J. J., et al., Origins of Ultralow Thermal Conductivity in 1-2-1-4 Quaternary Selenides. *J Mater Chem A* **2019**, *7*, 2589-2596.

27. Zhu, H. T., et al., Discovery of Tafesb-Based Half-Heuslers with High Thermoelectric Performance. *Nat Commu* **2019**, *10*, 270-270.

28. Xie, W. J.; Weidenkaff, A.; Tang, X. F.; Zhang, Q. J.; Poon, J.; Tritt, T. M., Recent Advances in Nanostructured Thermoelectric Half-Heusler Compounds. *Nanomaterials* **2012**, *2*, 379-412.

29. Yu, C.; Zhu, T. J.; Shi, R. Z.; Zhang, Y.; Zhao, X. B.; He, J., High-Performance Half-Heusler Thermoelectric Materials Hf$_{1-X}$ Zrxnisn$_{1-Y}$sb$_Y$ Prepared by Levitation Melting and Spark Plasma Sintering. *Acta Mater* **2009**, *57*, 2757-2764.

30. Culp, S. R.; Simonson, J. W.; Poon, S. J.; Ponnambalam, V.; Edwards, J.; Tritt, T. M., (Zr,Hf)Co(Sb,Sn) Half-Heusler Phases as High-Temperature ( > 700 ℃ ) P-Type Thermoelectric Materials. *Appl Phys Lett* **2008**, *93*, 022105-022105.

31. Zhu, T.; Huhn, W. P.; Wessler, G. C.; Shin, D.; Saparov, B.; Mitzi, D. B.; Blum, V., I-2-Ii-Iv-Vi4 (I = Cu, Ag; Ii = Sr, Ba; Iv = Ge, Sn; Vi = S, Se): Chalcogenides for Thin-Film Photovoltaics. *Chem Mater* **2017**, *29*, 7868-7879.

32. Sun, J. P.; Wessler, G. C. M.; Wang, T. L.; Zhu, T.; Blum, V.; Mitzi, D. B., Structural Tolerance Factor Approach to Defect-Resistant I-2-Ii-Iv-X-4 Semiconductor Design. *Chem Mater* **2020**, *32*, 1636-1649.

33. Nian, L. Y.; Wu, K.; He, G. J.; Yang, Z. H.; Pan, S. L., Effect of Element Substitution on Structural Transformation and Optical Performances in I$^2$bam$^{iv}$q$^4$ (I = Li, Na, Cu, and Ag; M-Iv = Si, Ge, and Sn; Q = S and Se). *Inorg Chem* **2018**, *57*, 3434-3442.

34. Chen, S. Y.; Walsh, A.; Gong, X. G.; Wei, S. H., Classification of Lattice Defects in the Kesterite Cu$_2$znsns$_4$ and Cu$_2$znsnse$_4$ Earth-Abundant Solar Cell Absorbers. *Adv Mater* **2013**, *25*, 1522-1539.

35. Wang, W.; Winkler, M. T.; Gunawan, O.; Gokmen, T.; Todorov, T. K.; Zhu, Y.; Mitzi, D. B., Device Characteristics of Cztsse Thin-Film Solar Cells with 12.6% Efficiency. *Adv Energy Mater* **2014**, *4*, 1301465-1301465.

36. Schorr, S.; Hoebler, H. J.; Tovar, M., A Neutron Diffraction Study of the Stannite-Kesterite Solid Solution Series. *Eur J Mineral* **2007**, *19*, 65-73.

37. Perdew, J. P.; Burke, K.; Ernzerhof, M., Generalized Gradient Approximation Made Simple. *Phys Rev Lett* **1996**, *77*, 3865-3868.

38. Tran, F.; Blaha, P., Accurate Band Gaps of Semiconductors and Insulators with a Semilocal Exchange-Correlation Potential. *Phys Rev Lett* **2009**, *102*, 226401-226401.

39. Schwarz, K.; Blaha, P.; Madsen, G. K. H., Electronic Structure Calculations of Solids Using the Wien2k Package for Material Sciences. *Comput Phys Commun* **2002**, *147*, 71-76.

40. Madsen, G. K. H.; Singh, D. J., Boltztrap. A Code for Calculating Band-Structure Dependent Quantities. *Comput Phys Commun* **2006**, *175*, 67-71.

41. Zhou, F.; Nielson, W.; Xia, Y.; Ozolins, V., Lattice Anharmonicity and Thermal Conductivity from Compressive Sensing of First-Principles Calculations. *Phys Rev Lett* **2014**, *113*, 185501-185501.

42. Zhou, F.; Nielson, W.; Xia, Y.; Ozolins, V., Compressive Sensing Lattice Dynamics. I. General Formalism. *Phys Rev B* **2019**, *100*, 184308-184308.

43. Li, W.; Carrete, J.; Katcho, N. A.; Mingo, N., Shengbte: A Solver of the Boltzmann Transport Equation for Phonons. *Comput Phys Commun* **2014**, *185*, 1747-1758.

44. Madsen, G. K. H., Automated Search for New Thermoelectric Materials: The Case of Liznsb. *J Am Chem Soc* **2006**, *128*, 12140-12146.

45. Bhattacharya, S.; Madsen, G. K. H., A Novel P-Type Half-Heusler from High-Throughput Transport and Defect Calculations. *J Mater Chem C* **2016**, *4*, 11261-11268.

46. Liu, M. L.; Huang, F. Q.; Chen, L. D.; Chen, I. W., A Wide-Band-Gap P-Type Thermoelectric Material Based on Quaternary Chalcogenides of $Cu_2ZnSnQ_4$ (Q=S,Se). *Appl Phys Lett* **2009**, *94*, 202103-202103.

47. Takabatake, T.; Suekuni, K.; Nakayama, T.; Kaneshita, E., Phonon-Glass Electron-Crystal Thermoelectric Clathrates: Experiments and Theory. *Rev Mod Phys* **2014**, *86*, 669-716.

48. Xu, N.; Xu, Y.; Zhu, J., Topological Insulators for Thermoelectrics. *npj Quantum Materials* **2017**, *2*, 51.

49. Wei, K.; Neu, J. N.; Lai, Y.; Chen, K.-W.; Hobbis, D.; Nolas, G. S.; Graf, D. E.; Siegrist, T.; Baumbach, R. E., Enhanced Thermoelectric Performance of Heavy-Fermion Compounds $YbTm_2Zn_{20}$ (Tm = Co, Rh, Ir) at Low Temperatures. *Science Advances* **2019**, *5*, 6183.

50. Gibbs, Z. M.; Ricci, F.; Li, G. D.; Zhu, H.; Persson, K.; Ceder, G.; Hautier, G.; Jain, A.; Snyder, G. J., Effective Mass and Fermi Surface Complexity Factor from Ab Initio Band Structure Calculations. *Npj Computational Materials* **2017**, *3*, 8.

51. Zhao, L. D., et al., Ultrahigh Power Factor and Thermoelectric Performance in Hole-Doped Single-Crystal Snse. *Science* **2016**, *351*, 141-144.

52. Xin, J. Z.; Tang, Y. L.; Liu, Y. T.; Zhao, X. B.; Pan, H. G.; Zhu, T. J., Valleytronics in Thermoelectric Materials. *npj Quantum Materials* **2018**, *3*, 9.

53. Singh, D. J.; Mazin, I. I., Calculated Thermoelectric Properties of La-Filled Skutterudites. *Phys Rev B* **1997**, *56*, R1650-R1653.

54. Xiao, Y.; Zhao, L. D., Charge and Phonon Transport in Pbte-Based Thermoelectric Materials. *npj Quantum Materials* **2018**, *3*, 55.

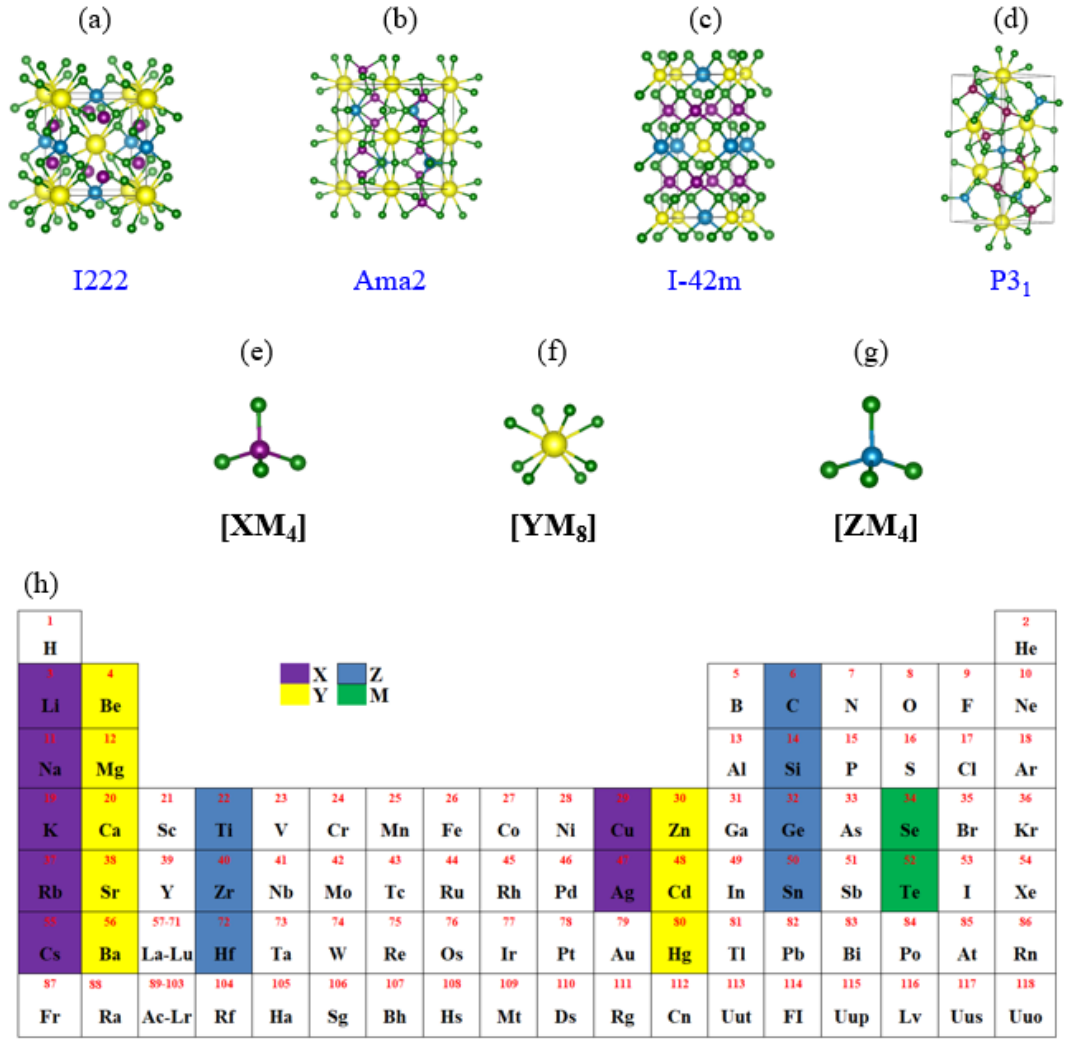

**Figure 1.** Schematic illustrations of the lattice structures for the $X_2YZM_4$ quaternary compounds with the space group: (a) I222, (b) Ama2, (c) I-42m, and (d) P3$_1$. In these structures, several core polyhedron units essential for determining the electronic and phonon transports are shown: (e) [XM$_4$], (f) [YM$_8$], and (g) [ZM$_4$]. (h) The preferred elements for X, Y, and Z are marked while M = Se and Te.

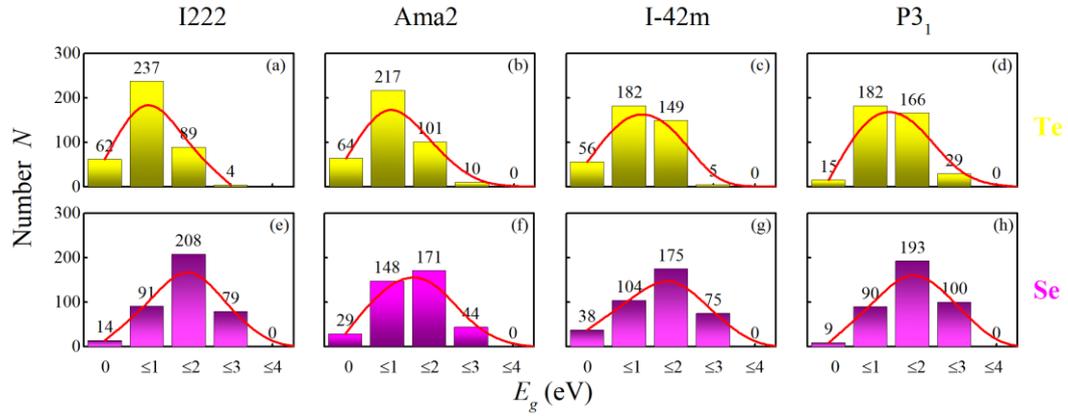

**Figure 2.** The histograms for illustrating the number of compounds $N$ .vs. PBE-bandgap $E_g$ for (a) ~ (d) the Te-based compounds and (e) ~ (h) the Se-based compounds in the I222, Ama2, I-42m and P3$_1$ phases respectively. The bandgap data are determined from the calculations based on the PBE scheme and they usually are slightly under-estimated. The more accurate data of bandgap are obtained from the calculations based on the mBJ scheme.

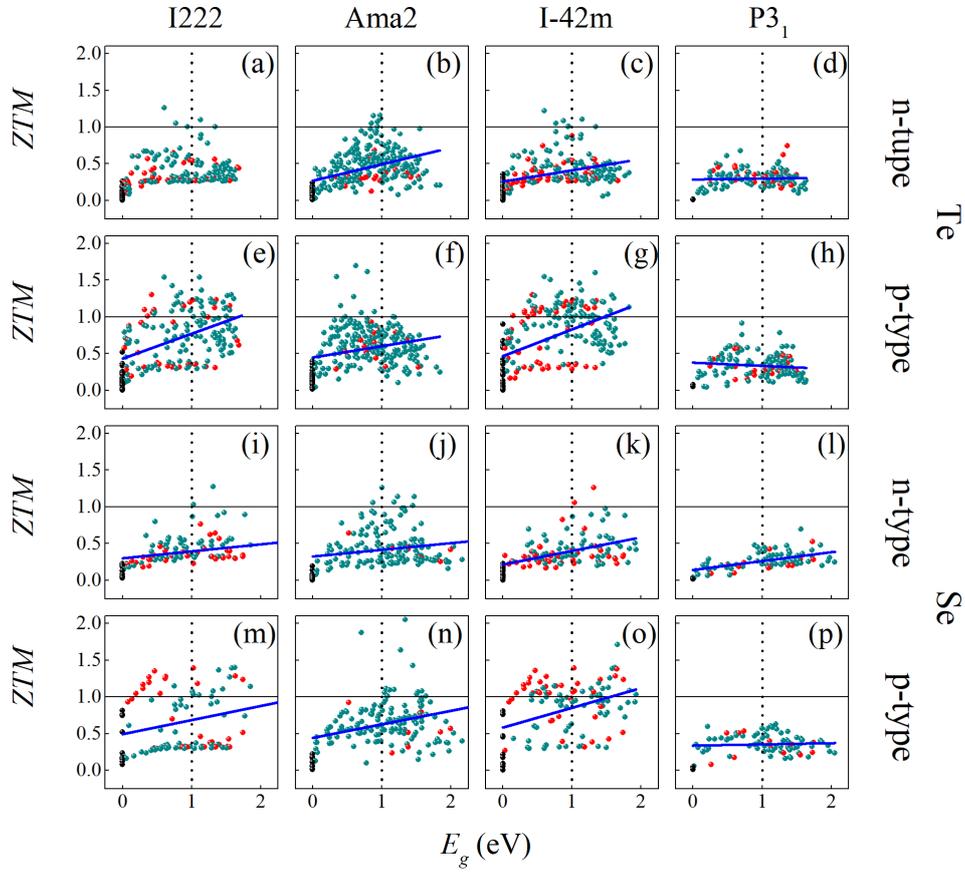

**Figure 3.** The collective plots of the *ZTM* .vs. $E_g$ data for (a)-(d) the *n*-type and (c)-(h) *p*-type Te-based compounds, and (i)-(l) the *n*-type and (m)-(p) *p*-type Se-based compounds in the I222, Ama2, I-42m, and P3$_1$ phases respectively. Each dot represents the data for one compound, and the bandgap data are determined from the calculations based on the mBJ scheme. The blue lines are the best fitting results. The black and red/olive spheres in each plot respectively represent the zero-bandgap (metal), indirect bandgap, and direct bandgap.

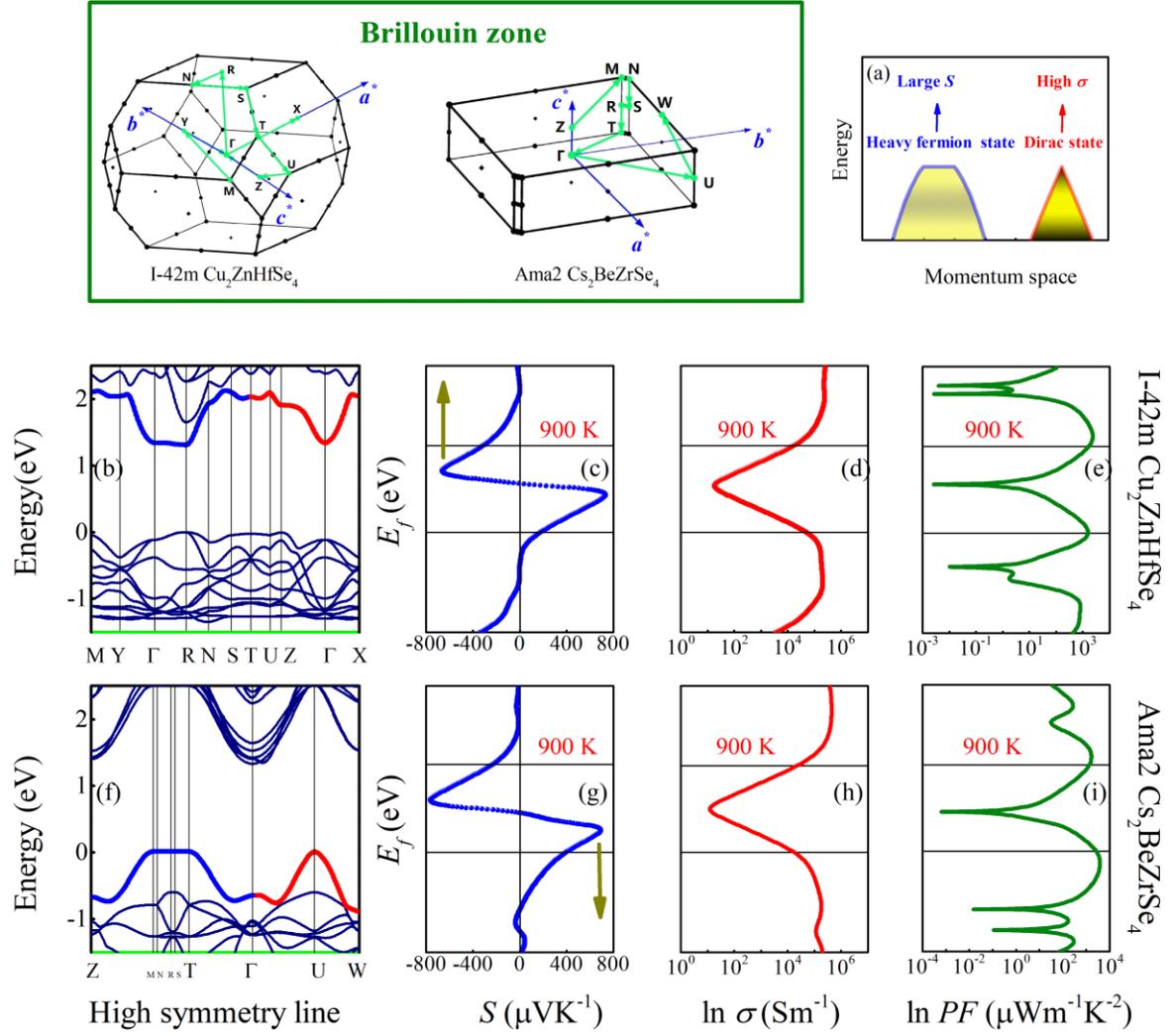

**Figure 4**. (a) Ideal valley band model for CDHF systems, band structures, $E_f$-dependent $S$, $\sigma$ and $PF$ for (b)-(e) I-42m $Cu_2ZnHfSe_4$ and (f)-(i) Ama2 $Cs_2BeZrSe_4$ (the left to the right). The high symmetry lines [the green abscissas in (b) and (f)] for band structure correspond to the green polylines in Brillouin zone plot shown on the top.

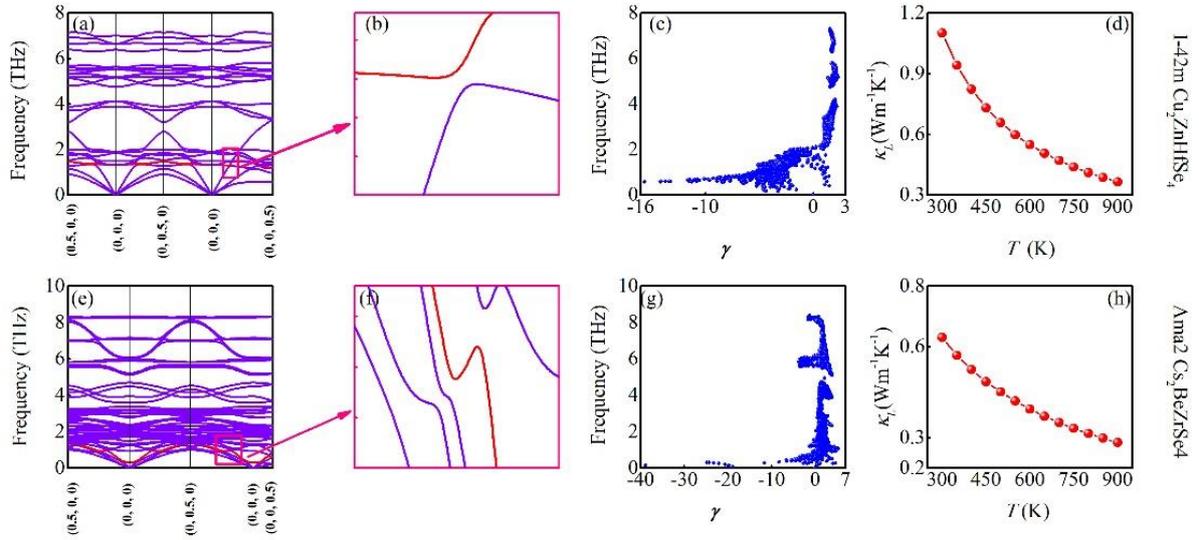

**Figure 5.** The phonon spectrum, enlarged avoided crossing, frequency-dependent $\gamma$ and $T$-dependent $\kappa_L$ for (a)-(d) I-42m $Cu_2ZnHfSe_4$ and (e)-(h) Ama2 $Cs_2BeZrSe_4$. The pink rectangles in (a) and (e) mark avoided crossings that are enlarged in (b) and (f) respectively, and the red lines in (a), (b), (e) and (f) are on behalf of the lowest-lying optical (LLO) branches.

TOC Graphic

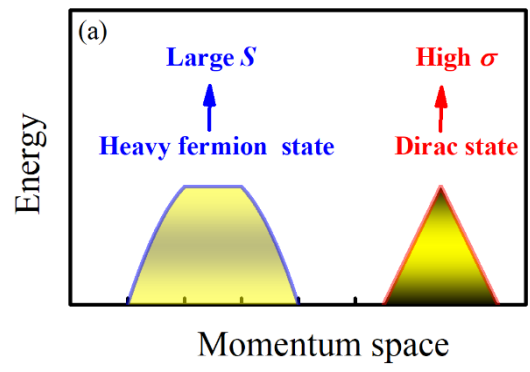